\documentstyle[12pt]{article}
\begin{document}
\newcommand{\beq}{\begin{equation}}
\newcommand{\eeq}[1]{\label{#1} \end{equation}}
\newcommand{\beqar}{\begin{eqnarray}}
\newcommand{\eeqar}[1]{\label{#1} \end{eqnarray}}
\def\p{ \partial }

\hfill KSUCNR-012-93 revised
\vspace{1truecm}
\begin{center}
\begin{large}
{\bf Classical Lagrangian Model of the Pauli Principle}
\end{large}
\vskip .5 truecm
{\bf John J. Neumann and George Fai}\\
\vskip .2 truecm
Center for Nuclear Research\\
Department of Physics\\
Kent State University, Kent, OH 44242\\
\vskip 2 truecm
{\bf ABSTRACT}\\
\end{center}
\begin{quotation}
A classical Lagrangian model of the Pauli potential
is introduced. It is shown that the
kinematic kinetic energy ($\sum \frac{1}{2} m v^2$)
in the model approximately reproduces the energy of a free Fermi gas
at low temperatures and at densities
relevant in nuclear collisions with moderate beam energies.
Differences between canonical and kinematic quantities are pointed out.
The Pauli potential can be used in transport simulations.
\end{quotation}
\vspace{.6truecm}
\setcounter{section}{0}
\setcounter{equation}{0}

\newpage
Nuclear collisions at beam energies on the order of 100 MeV/nucleon
offer a unique testing ground for the dynamics of
many-fermion systems. After pioneering studies at the Berkeley
Bevalac, the SIS-18 accelerator in Darmstadt has now become the
main theater for the experimental investigation of the rich variety
of phenomena displayed in these collisions.

Semiclassical approximations proved to be useful in modeling nuclear
collisions at these energies. The resulting
transport codes, which are based on the Boltzmann equation,
are at present the main tools used to extract
physical information from the data.
Although these models are popular and successful, most of them
make a large number of simplifying assumptions and face
difficulties with some aspects of the physics as well as with the
numerical implementation. Many of the difficulties can be traced back to
the problem of enforcing the restrictions on the occupancy of states
implied by the Fermi statistics of the nucleons.

An alternative approach is to seek a classical many-body description
simulating the effect of the Pauli principle. This can be accomplished
by a potential that keeps the nucleons apart in phase space
(`Pauli potential'). Such a classical model was first proposed by Wilets
and collaborators.\cite{wilets77}
It has been demonstrated
that many features of the Fermi gas can approximately be
reproduced with a suitably chosen Pauli potential.\cite{dorso87}
The Pauli potential has been incorporated in computational simulations
of nuclear collisions.\cite{wilets78,boal88,peilert92} One major advantage
of classical dynamical models is that once the model is defined,
no further approximations are necessary.

An important consequence of the introduction of the Pauli potential
is that differences appear between the canonical and the kinematic
quantities. This complication can be ignored as long as attention is
focused only on the canonical momenta.
Here we argue that one needs to keep track of the differences
between canonical and kinematic quantities, and that the latter
determine several physical observables of interest.
In particular, thermodynamic quantities will be sensitive to the
kinematic momenta. (Note that there are analogous
differences between the canonical and kinematic momenta
in the fermionic molecular dynamics
model developed by Feldmeier.\cite{feldmeier90})
The classical simulations mentioned above have all
been carried out in the Hamiltonian framework, where
the natural variables are the canonical rather than the kinematic momenta.

Motivated by the above arguments, we study a classical Lagrangian
model of the `Pauli potential,' which can be used in collision
simulations. The fermion nature of the nucleons, which is
the most important quantum feature in nuclear collisions,
should be retained in these simulations even if
other quantum corrections are neglected. In common with earlier work,
we aim at a model which implements
the `Pauli principle' in a classical framework (by this we mean
that identical nucleons must be kept apart in the phase space).
However, we depart from earlier studies
by presenting calculations in the Lagrangian formalism.

Formally, the Hamiltonian for any physical problem must be constructed
via the Lagrangian.\cite{Goldstein80} The Lagrangian is a
function of $n$ generalized coordinates $q_i$, their time-derivatives,
${\dot q}_i$, and possibly the time. (We will only deal with
Lagrangians $L(\ldots,q_i,{\dot q}_i,\ldots)$
without explicit time-dependence in the following.)
The Hamiltonian $H$ is obtained by introducing the canonical momenta,
$p_i = \p L/\p {\dot q}_i \,$, then constructing\\
$H = \sum {\dot q}_i p_i - L \,$, and expressing it
as a function of $q_i$ and $p_i$ only.

Pauli potentials have been defined in the past as (canonical)
momentum-dependent potentials for inclusion in the Hamiltonian
formalism.\cite{wilets77,wilets78,dorso87,boal88} We call this class of
potentials Hamiltonian Pauli potentials. As an example,
consider the Hamiltonian Pauli potential between two nucleons of relative
separation $x$ and relative {\em canonical} momentum $p$
(in one dimension),
\beq
V(x,p) = V_0 \exp(-s^2/2)   \,\, ,
\eeq{randruppot}
where $V_0$ is a constant $(V_0 > 0)$,
and the phase-space separation $s$ is given as
\beq
s = \sqrt{(x/x_0)^2+(p/p_0)^2} \,\,  ,
\eeq{pssep}
with appropriate length- and momentum-scales
$x_0$ and $p_0$, respectively.\cite{dorso87} To see
to what extent $p$ is different from the kinematic momentum, $\mu{\dot x}$,
one may use the canonical equation
\beq
{\dot x} = \frac{\p H}{\p p} = \frac{p}{\mu}\left(1-\frac{\mu V_0}{p_0^2}
		e^{-s^2/2}\right) \,\, ,
\eeq{caneq}
where $\mu$ is the reduced mass. In other words,
the velocity is decreased by the presence of the Pauli
potential. The effect is negligible for nucleons far apart in phase space,
but leads to a `crystallized' ground state if the same potential is used
in the many-body problem representing a nucleus.\cite{dorso87}
In the many-body ground state the nucleons have finite (canonical)
momenta, but vanishing velocities.

A Pauli potential similar to the one given in eq. (\ref{randruppot})
has been arrived at based on a consideration of the evolution of two
Gaussian wave packets,\cite{boal88} while a Hamiltonian Pauli potential
with quite different scaling properties is used
by Wilets and collaborators (also in atomic processes).\cite{wilets91}

The one-dimensional two-body problem has recently been examined with
a simplified
version of the Pauli potential, where it was possible
to carry out all calculations in both the Lagrangian and Hamiltonian
formalisms.\cite{nf93}
The difference between the canonical and the kinematic quantities
has been demonstrated in this simplified framework.

Here we employ a more realistic Lagrangian Pauli potential
inspired by the Hamiltonian form (\ref{randruppot}). We take
\beq
V(x,{\dot x}) = V_0 \exp(-t^2/2)   \,\, ,
\eeq{Lpot}
where the separation $t$ in $(x,\dot{x})$ space is given by
\beq
t = \sqrt{(x/x_0)^2+({\dot x}/v_0)^2} \,\,  ,
\eeq{xdotpsep}
with scale parameters $x_0$ and $v_0$. For the two-body problem
defined by this Lagrangian Pauli potential,
\beq
p=\frac{\p L}{\p {\dot x}}=
	\left(1+\frac{V_0}{\mu v_0^2} e^{-t^2/2}\right) \mu {\dot x}  \,\, .
\eeq{canmom}
It is worth noting that $\mu{\dot x}$ is the relevant
quantity in nucleon-nucleon collisions, cross sections, and with respect to
the thermodynamics of the system.

We wish to investigate the finite-temperature
behavior of a system of particles interacting via the
Lagrangian Pauli potential (\ref{Lpot}). The three-dimensional
many-body Lagrangian takes the form
\beq
L = \frac{m_N}{2} \sum_i ({\dot x}_i^2+{\dot y}_i^2+{\dot z}_i^2)
- \frac{V_0}{2} \sum_{i \neq j}
\exp\left[-\left\{
\left({r_{ij} \over r}\right)^2+\left({v_{ij} \over v}\right)^2\right\}\right]
\,\,  ,
\eeq{reallag}
with relative coordinates
$r_{ij} = \sqrt{(x_i-x_j)^2+(y_i-y_j)^2+(z_i-z_j)^2} \,\,$, and
relative velocities
$v_{ij} = \sqrt{({\dot x}_i-{\dot x}_j)^2+({\dot y}_i-{\dot y}_j)^2
+({\dot z}_i-{\dot z}_j)^2} \,\,$.
As before, $V_0$ is the overall strength, $r$ and $v$ are
scale parameters, and $m_N$ stands for the nucleon mass.

In what follows, we focus on the behavior of the many-body system at
low temperatures (up to $\approx 5$ MeV) and densities corresponding to
deviations from standard nuclear matter density that are
not too large. The latter requirement
means that densities around 0.04 fm$^{-3}$ will be considered for
a given type of nucleon with a given spin orientation.

Thermal averaging is carried out in the phase space
defined by canonically conjugate variables.
Schematically, for an observable $A(x,p)$,
\beq
{\overline A} = \frac{\int dx dp e^{-\beta H} A(x,p)}
                     {\int dx dp e^{-\beta H}} \,\,\,\, ,
\eeq{av}
where the integrals are over the phase space, and
$\beta$ is the inverse temperature. We emphasize that
$p$ in eq. (\ref{av}) stands for the {\em canonical} momentum.
It is therefore necessary to first map the $(x,\dot{x})$ space
to the $(x,p)$ space in the present formalism before applying the
Metropolis algorithm\cite{metropolis53} to calculate averages.
The mapping  in the $\dot{x} \Rightarrow p$
direction poses no technical problem with our Lagrangian Pauli potential,
so each trial step is made first in $(x, \dot x)$ space and then the
corresponding step in $(x,p)$ space is measured. This measurement is used to
rescale the $(x, \dot x)$ step so that we get a uniform walk through
$(x,p)$ space. In order to maintain full control over the resulting step size
in the $p$ subspace, it is necessary to hold the $x$'s constant while changing
the $\dot x$'s. For this reason we split the calculation into two parts:
a standard Monte Carlo sampling of $x$-configurations, and a Metropolis
sampling over the $p$-subspace (``$p$-Metropolis procedure'')
for each $x$-configuration.

Several (typically ten) $x$-configurations are chosen for each temperature
by filling a specified volume (periodic boundary conditions)
with $N$ nucleons at random. We have chosen
$N=60$ for the illustrative results presented here. We examined the dependence
of our calculations on the number of nucleons and found small
surface effects to be still present for $N=60$. However, since
the required CPU time is approximately quadratic in $N$, we
carried out most of our calculations with this value. The surface effects can
be decreased by decreasing the range of the Pauli potential in
coordinate space.

The nucleons are held at fixed positions while their momenta
are changed during the $p$-Metropolis procedure.
The results for the different $x$-configurations are then averaged with
equal weights, and their spread around the average divided by the
square root of the number of $x$-configurations provides an estimate of
the statistical uncertainty in the result for the given temperature.

To achieve fast equilibration, the step size is continually adjusted
at the beginning of the $p$-Metropolis procedure,
so that about half of the attempted steps are accepted (or rejected).
We assume that equilibrium has been sufficiently approached when the
procedure yields the first extremum of $\overline H$. Averaging is
started at this point with a constant step size in momentum space.

Specifically, we put $A = E_{kin}/N$ in eq. (\ref{av}), where by $E_{kin}$
we mean the {\em kinematic} kinetic energy,
\beq
E_{kin} = \sum_i \frac{1}{2} m_N v_i^2 = \frac{m_N}{2}
\sum_i ({\dot x}_i^2 + {\dot y}_i^2 + {\dot z}_i^2)    \,\,   .
\eeq{Ekin}
We study the dependence of this quantity on the temperature and density,
since (as mentioned above) the kinematic quantities are most important for
the purposes of simulations and thermodynamics. We therefore wish to judge
the success of our model by the extent to which (\ref{Ekin}) reproduces the
kinetic energy of a Fermi gas (which, in the case of a free Fermi gas,
is identical to the total energy).

The temperature dependence of the results is displayed on Fig. 1
for the parameters $V_0=1.17$ GeV, $r=1.5$ fm, and $v=0.163$c.
The density has been fixed at the value
corresponding to standard nuclear matter density.
The calculations are represented by the points with their
statistical uncertainties at different temperatures.
The dashed line corresponds to a quadratic least-square fit through these
points with reduced $\chi^2 \approx 0.81$. (A linear least-square fit
would yield reduced $\chi^2 \approx 1.23$.) Also shown (solid line) is
the temperature dependence of the energy per particle of an ideal Fermi gas
in the low-temperature approximation
(through second order in the ratio of the
temperature to the Fermi energy, $T/\epsilon_F$). The three parameters
of the Pauli potential can be used to approximately reproduce the
ground-state energy of the Fermi gas, $\frac{3}{5}\epsilon_F$, and
can be traded against each other to some extent
to obtain fits of similar quality. We chose a strong Pauli potential
with relatively small ranges.
The calculated results are not inconsistent with the desired
behavior in the temperature range of interest, but the fit
overestimates the specific heat.

Fig. 2 shows the dependence of the kinematic kinetic energy per nucleon
on the density at a temperature of $0.5$ MeV
for the same values of the parameters.
It is seen that the desired Fermi-gas behavior (solid curve) is approximately
reproduced. Numerically, the agreement can be judged sufficiently close
up to densities corresponding to about 1.5 standard
nuclear matter density.

Fig. 3 displays an example of the distribution function in
the kinematic kinetic energy, $\sum \frac{1}{2} m v^2$ at a
temperature $T=5$ MeV in our model. While the histogram
(model) reproduces the Fermi gas (dashed line) only approximately, we
can see that the occupation of the low energies is limited, in contrast
to the expectation for a classical system without the Pauli potential.
The distributions at other temperatures have qualitatively similar
features.

In the present work we have calculated the temperature and density dependence
of the kinematic kinetic energy of a classical system of particles under
the influence of a Pauli potential.
The precise meaning of the Pauli principle in the classical framework
used here and in standard Hamiltonian simulations is of course open to
interpretation to some degree. It is physically clear, however, that
the Pauli principle keeps identical nucleons apart in phase space.
Such a requirement
can be satisfied with both Lagrangian and Hamiltonian Pauli potentials.
Here we used this freedom of implementation to introduce a
Lagrangian Pauli potential. The kinematic kinetic energy in the model
approximately reproduces the energy of a free Fermi
gas at low temperature and at densities relevant in nuclear collisions
with moderate beam energies. We emphasized the role of the kinematic
quantities (as opposed to canonical momenta) for thermodynamic
purposes. These kinematic quantities can be calculated in the model.
We foresee applications of this
Pauli potential to situations wherein fragmentation or evaporation
at low temperatures plays an important role.
\vspace{3truemm}

Helpful discussions with Dr. F. Gulminelli are gratefully acknowledged.
This work was supported in part by the
U. S. DOE under Grant DE-FG02-86ER40251.

\newpage
\section*{Figure Captions}

\begin{description}
\item[Fig. 1:] The average kinematic kinetic energy per nucleon as a function
of the temperature at the density $\rho = 0.04$ fm$^{-3}$. The dashed line
corresponds to a quadratic least-square fit as explained in the text.
The solid line represents an ideal Fermi gas in the low-temperature
approximation.
\item[Fig. 2:] The average kinematic kinetic energy per nucleon as a function
of the density at temperature $T = 0.5$ MeV. The dashed line is a fit through
the data. The solid line displays the Fermi-gas result.
\item[Fig. 3:] The distribution function in terms of the kinematic kinetic
energy at temperature $T = 5$ MeV. The histogram represents the model,
the dashed line corresponds to an ideal Fermi gas.
\end{description}
\end{document}